\documentclass{eas}
\usepackage{graphicx}
\newcommand{\Eq}[1]{Eq.~(\ref{#1})}

\newcommand{\ea}{et~al.~}                            

\newcommand{\astroph}[1]{\mbox{\texttt{astro-ph/#1}}}
\begin{document}

\title{A correlation between spin parameter and dark matter halo mass} 
\runningtitle{Spin parameter and halo mass}
\author{Alexander Knebe}\address{Astrophysikalisches Institut Potsdam, An der Sternwarte 16, 14482 Potsdam, Germany}
\author{Chris Power}\address{Department of Physics and Astronomy, University of Leicester, University Road, LE1 7RH, UK}
%
%
\begin{abstract}
  Using a set of high-resolution dark matter only cosmological
  simulations we found a correlation between the dark matter halo mass
  $M$ and its spin parameter $\lambda$ for objects forming at
  redshifts $z > 10$: the spin parameter decreases with
  increasing mass.  However, halos forming at later times do not
  exhibit such a strong correlation, in agreement with the findings of
  previous studies.

  While we presented such a correlation in a previous study using the
  Bullock et al. (2001) spin parameter defintion we now defer to the
  classical definition showing that the results are independent of the
  definition.

\end{abstract}
\maketitle
\section{Introduction}
The physical mechanism by which galaxies acquire their angular momentum
is an important problem that has been the subject of investigation for
nearly sixty years (Hoyle 1949). This reflects the fundamental role played
by angular momentum of galactic material in defining the size and 
shapes of galaxies (e.g. Fall \& Efstathiou 1981). Yet despite its
physical significance, a precise and accurate understanding of the
origin of galactic angular momentum remains one of the missing pieces 
in the galaxy formation puzzle.

A fundamental assumption in current galaxy formation models is that
galaxies form when gas cools and condenses within the potential wells
of dark matter halos (White \& Rees 1978). Consequently it is probable that 
the angular momentum of the galaxy will be linked to the angular
momentum of its dark matter halo (e.g. Fall \& Efstathiou 1980; Mo, Mao
\& White 1998; Zavala, Okamoto \& Frenk 2007). Within the context
of hierarchical structure formation models, the angular momentum growth 
of a dark matter proto-halo is driven by gravitational tidal torquing during 
the early stages (i.e. the linear regime) of its assembly. This ``Tidal Torque
Theory'' has been explored in detail; it is a well-developed analytic 
theory (e.g. Peebles 1969, Doroshkevich 1979, White 1984) and its 
predictions are in good agreement with the results of cosmological 
$N$-body simulations (e.g. Barnes \& Efstathiou 1987; Warren~\ea 1992; 
Sugerman, Summers \& Kamionkowski 2000; Porciani, Dekel \& Hoffman
2002). However, once the proto-halo has passed through maximum
expansion and the collapse has become non-linear, tidal torquing
no longer provides an adequate description of the evolution of the 
angular momentum (White 1984), which tends to decrease with time. 
During this phase it is likely that merger and accretion 
events play an increasingly important role in determining both the
magnitude and direction of the angular momentum of a galaxy (e.g. Bailin \&
Steinmetz 2005). Indeed, a number of studies have argued that mergers
and accretion events are the primary determinants of the angular
momenta of galaxies at the present day (Gardner 2001; Maller, Dekel~\&
Somerville 2002; Vitvitska~\ea 2002).

It is common practice to quantify the angular momentum of a dark matter 
halo by the dimensionless ``classical'' spin parameter (Peebles 1969),
\begin{equation}
  \label{eq:lambda}
  \lambda = \frac{J \sqrt{|E|}}{GM^{5/2}},
\end{equation}
where $J$ is the magnitude of the angular momentum of material within
the virial radius, $M$ is the virial mass, and $E$ is the total energy 
of the system. It has been shown that halos that have suffered a recent 
major merger will tend to have a higher spin parameter $\lambda$ than 
the average (e.g. Hetznecker \& Burkert 2006; Power, Knebe \& Knollmann 
2008). Therefore one could argue that within the framework of hierarchical
structure formation that higher mass halos should have larger spin 
parameters \emph{on average} than less massive systems because they 
have assembled a larger fraction of their mass (by merging) more
recently. 

However, if we consider only halos in virial equilibrium, should we
expect to see a correlation between halo mass and spin? One might
na\"ively expect that more massive systems will have had their maximum
expansion more recently and so these systems will have been tidally
torqued for longer than systems that had their maximum expansion at
earlier times. This suggests that spin should \emph{increase} with
timing of maximum expansion and therefore halo mass. However, one
finds at best a weak correlation between mass and spin for equilibrium
halos at $z$=0 (e.g. Cole~\& Lacey 1996; Bett et al. 2007), and the
correlation is for spin to \emph{decrease} with increasing halo mass,
contrary to our na\"ive expectation.

\section{The Data}
For the simulations presented here we have adopted the cosmology as
given by Spergel~\ea (2003) ($\Omega_0=0.3$, $\Omega_{\Lambda}=0.7$,
$\sigma_8=0.9$, and $H_0=70$km/sec/Mpc). Each run employed $N=256^3$
particles and differed in simulation box-size $L_{\rm box}$, which
leads to the particle mass $m_p$ differing between runs --
$m_p=\rho_{\rm crit} \Omega_0 (L_{\rm box}/N)^3$, where $\rho_{\rm
  crit}=3H_0^2/8\pi\,G$. This allows us to probe a range of halo
masses at redshift $z$=10. Halos in all runs have been identified
using the MPI version of the \texttt{AHF} halo
finder\footnote{\texttt{AHF} is already freely available from
  \texttt{http://www.aip.de/People/aknebe}}
(\texttt{AMIGA}'s-Halo-Finder), which is based on the \texttt{MHF}
halo finder of Gill, Knebe \& Gibson (2004). Because we wish to
examine the spin distribution of equilibrium halos, it is important to
account for unrelaxed systems when investigating correlations between
spin and halo mass. For an eloborate discussion of our relaxation
criterion as well as more details about the simulations and halos we
refer the reader to the original paper (Knebe~\& Power 2008).

\section{Correlation between Spin and Mass}
While in Knebe~\& Power (2008) we used the spin-parameter
$\lambda'=J/\sqrt{2}MVR$ as defined by Bullock et al. (2001) we show
results in this contribution based upon the classical spin-parameter
definition as given by \Eq{eq:lambda}. Therefore, the work presented
here can be seen complemetary to the original study of Knebe~\& Power
(2008) and confirms that the findings are independent of the
spin-parameter definition.

\begin{figure} \label{fig:SpinMass}
\includegraphics[width=0.45\textwidth]{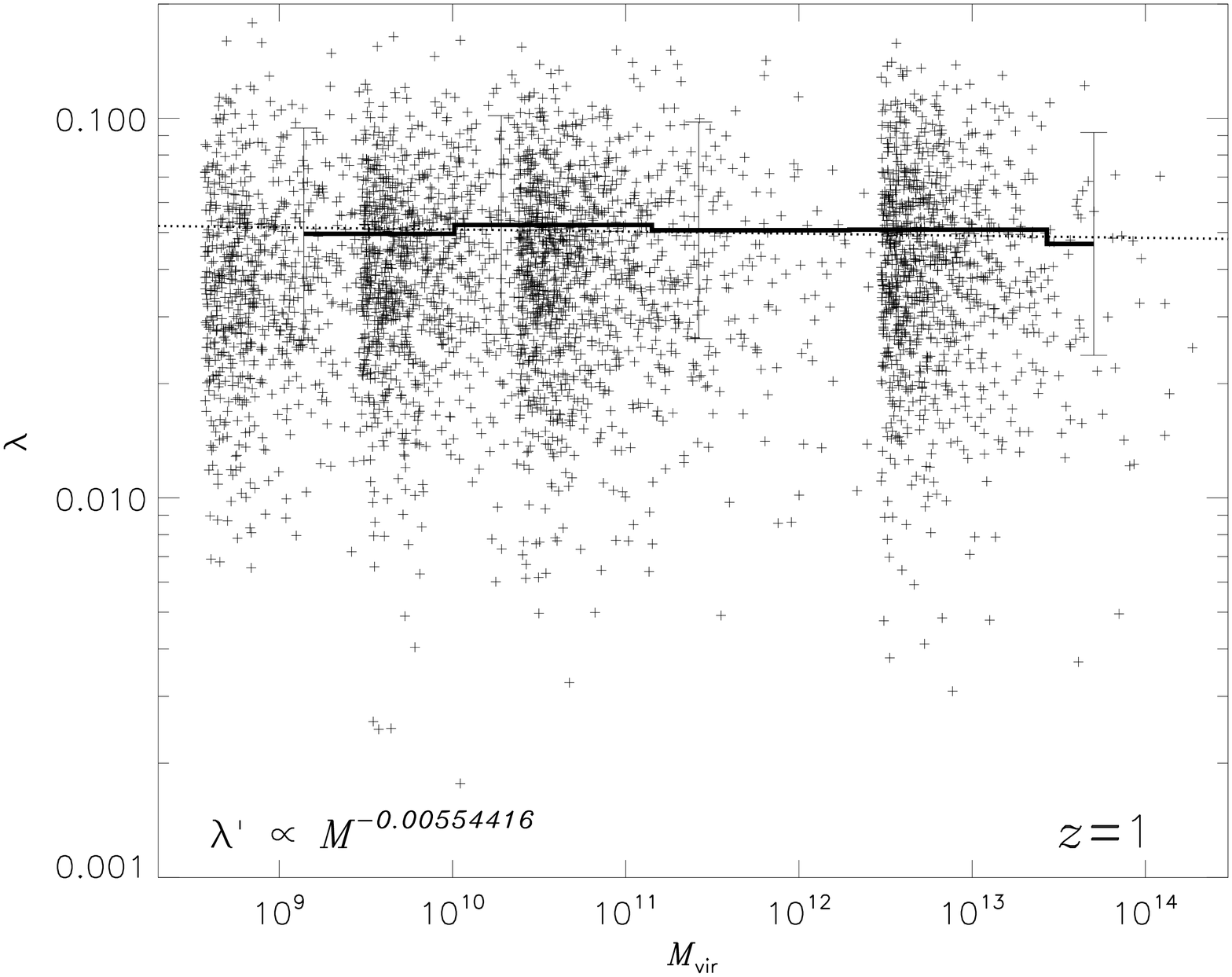} 
\qquad
\includegraphics[width=0.45\textwidth]{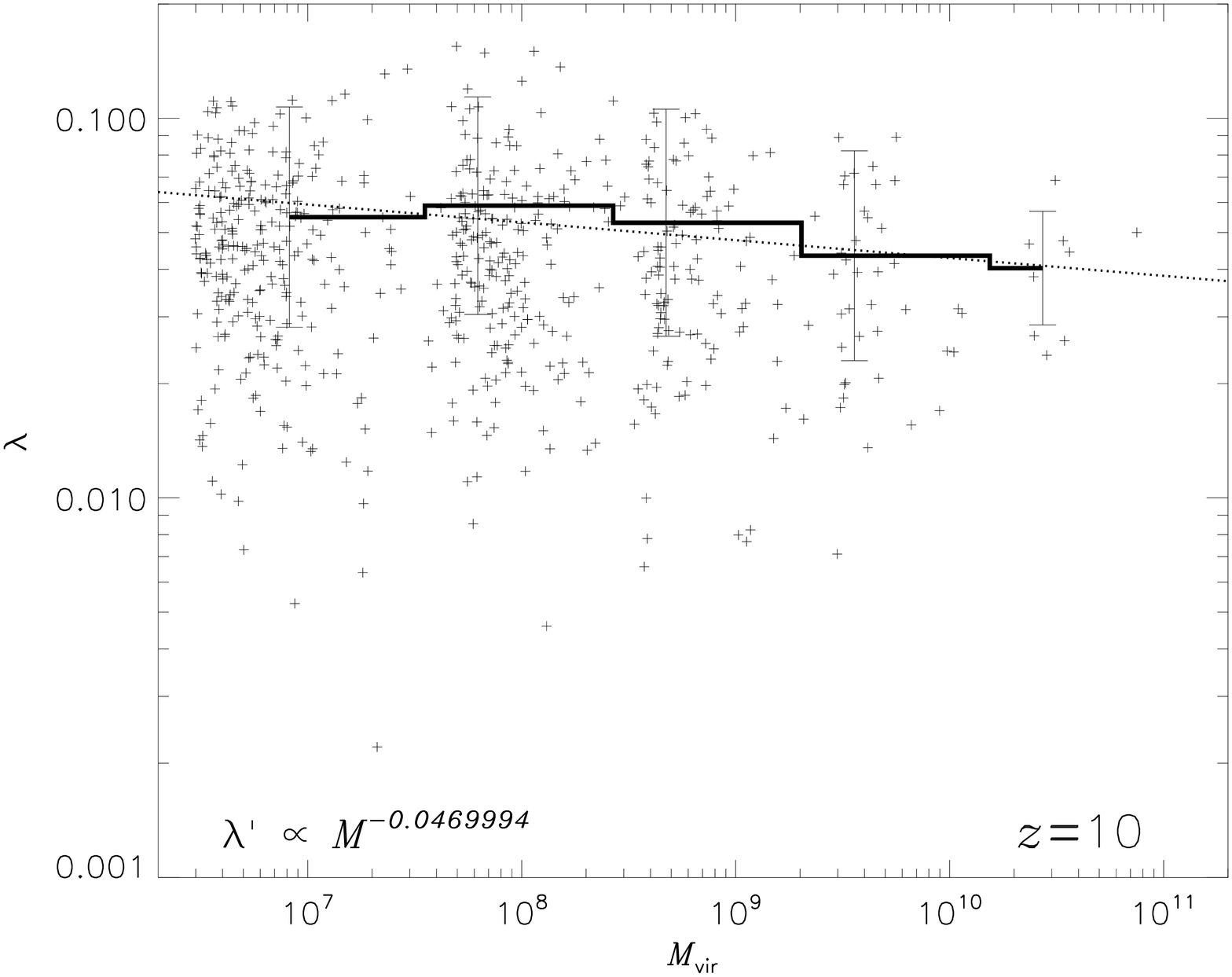} 
\caption{Correlation of spin parameter $\lambda$ with mass $M$.} 
\end{figure}

In Figure~\ref{fig:SpinMass}, we investigate the correlation between
halo spin $\lambda$ and mass $M$. The best fitting power-laws to these
histograms reveal that $\lambda \propto M^{\alpha}$ with

\begin{equation}\label{eq:logslopes}
\begin{array}{lcll}
\displaystyle \alpha  & = & -0.006 \pm 0.146 & \mbox{\rm ,\ for $z=1$ } \\
\displaystyle \alpha  & = & -0.047 \pm 0.168 & \mbox{\rm ,\ for $z=10$ } .\\ 
\end{array}
\end{equation}
This indicates that there is a \emph{weak} correlation at high
redshifts for spin to decrease with increasing mass, albeit stronger than the
one at $z$=1.

\section{Conclusions}
We have performed an investigation of the relation between virial mass
and dimensionless spin parameter for dark matter halos forming at high
redshifts $z > 10$ in a $\Lambda$CDM cosmology.  The result of our
study, which is based on a series of cosmological $N$-body simulations
in which box size was varied while keeping particle number fixed,
indicates that there is a \emph{weak} correlation between mass and
spin at $z$=10, such that the spin decreases with increasing mass. If
there is a correlation at $z$=1, we argue that it is significantly
weaker than the one we find at $z$=10; this is in qualitative
agreement with the findings of previous studies that focused on lower
redshifts (Maccio~\ea 2007, Shaw~\ea 2005, Lemson~\& Kauffmann 1999).

While we presented such a correlation in a previous study (Knebe~\&
Power 2008) using the Bullock et al. (2001) spin parameter defintion
we now defered to the classical definition showing that the results
are independent.


\end{document}